\documentclass[11pt]{article}

\usepackage{amsmath,fullpage}
\usepackage{xspace}
\usepackage{amssymb}
\usepackage{epsfig}

\newenvironment{Proof}[1][Proof.]{
	\par
	\noindent \textbf{#1}
}{
	\nobreak\leavevmode
	\hfill $\Box$\par\bigskip
}
\newtheorem{Thm}{Theorem}
\newtheorem{Lem}{Lemma}
\newtheorem{Cor}{Corollary}

\newtheorem{Def}{Definition}
\newtheorem{Fact}{Fact}

\newcommand{\defeq}{\stackrel{\mathsf{def}}{=}}

\newcommand\mcR{\mathcal{R}}

\mathchardef\mhyphen="2D

\newcommand{\disc}{{\mathsf{disc}}}
\newcommand{\sdisc}{{\mathsf{sdisc}}}
\newcommand{\tsdisc}{{\widetilde{\mathsf{sdisc}}}}

\newcommand\poly{{\mathsf{poly}}}
\newcommand\Ix{{\mathsf{Ix}}}
\newcommand\MajIx{{\mathsf{MajIx}}}
\newcommand\AppMP{{\mathsf{AppMP}}}
\newcommand\AppMPC{{\mathsf{AppMPC}}}
\newcommand\PAppMPC{{\mathsf{PAppMPC}}}
\newcommand\PAppMP{{\mathsf{PAppMP}}}
\newcommand\Disj{{\mathsf{Disj}}}
\newcommand\IP{{\mathsf{IP_2}}}
\newcommand\pdisc{{\mathsf{pdisc}}}
\newcommand\spdisc{{\mathsf{spdisc}}}

\newcommand{\suppress}[1]{}
\newcommand\COMMENT[1]{}

\begin{document}
\title{On Arthur Merlin Games in Communication Complexity
\thanks{
Research at the Centre
for Quantum Technologies is funded by the Singapore Ministry of Education
and the National Research Foundation.}}
\author{Hartmut Klauck\\Centre for Quantum Technologies (NUS)\\
 and School of Physical and Mathematical Sciences\\ Nanyang Technological University, Singapore.\\ Email: {\tt hklauck@gmail.com}  }
\date{}
\maketitle

\begin{abstract}
We show several results related to interactive proof modes of communication complexity. First we show lower bounds for the QMA-communication complexity of the functions Inner Product and Disjointness. We describe a general method to prove lower bounds for QMA-communication complexity, and show how one can 'transfer' hardness under an analogous measure in the query complexity model to the communication model using Sherstov's pattern matrix method.
Combining a result by Vereshchagin and the pattern matrix method we find a communication problem with AM-communication complexity $O(\log n)$, PP-communication complexity $\Omega(n^{1/3})$, and QMA-communication complexity $\Omega(n^{1/6})$. Hence in the world of communication complexity noninteractive quantum proof systems are not able to efficiently simulate co-nondeterminism or interaction. These results imply that the related questions in Turing machine complexity theory cannot be resolved by 'algebrizing' techniques. Finally we show that in MA-protocols there is an exponential gap between one-way protocols and two-way protocols (this refers to the interaction between Alice and Bob). This is in contrast to nondeterministic, AM-, and QMA-protocols, where one-way communication is essentially optimal.
\end{abstract}

\section{Introduction}

In their seminal 1986 paper on 'Complexity Classes in Communication Complexity' \cite{bfs:classes} Babai et al.~define, among a host of other classes, the communication complexity analogues of the interactive proof classes AM and MA. In this context, for instance, the {\em class} MA consists of all communication problems that have MA-communication complexity at most $\poly\log n$. The MA-communication complexity is the optimal complexity of a protocol solving a communication problem with the help of a prover Merlin, and two verifiers Alice and Bob (Alice and Bob each see only their part of the input, while Merlin sees the whole input, However, Merlin cannot be trusted). Merlin sends a proof followed by a discussion between Alice and Bob. See Section 2.1 for definitions.

While the Turing machine versions of AM and MA (capturing interactive proof systems with a constant number of rounds between the prover and verifier resp. noninteractive, but randomized proof systems \cite{babai1,babai2}) played a crucial role in the subsequent development of theoretical computer science, their communication complexity analogues were probably considered too esoteric a topic to merit much consideration. One of the few results about them is a 2003 lower bound of $\Omega(\sqrt n)$ for the MA-communication  complexity of the Disjointness problem $\Disj$ by this author \cite{klauck:thresh} ($\Disj$ is  co-NP complete in the world of communication complexity). The same paper also relates these complexity measures to the power of the rectangle (aka corruption) bound in communication complexity. In 2004, Raz and Shpilka \cite{raz:am2} proved that there is a problem, for which its quantum communication complexity is exponentially smaller than its MA-communication complexity, providing another MA-communication complexity lower bound, as well as a problem, for which its Quantum MA (short QMA-) communication complexity is exponentially smaller than both the quantum (without prover) and the MA-communication complexities. Raz and Shpilka left proving lower bounds for the QMA-communication complexity of any function as an open problem.

 Surprisingly, in 2008 Aaronson and Wigderson \cite{aaronson:alg} not only showed that the mentioned lower bound for $\Disj$ is basically tight, but also gave a new incentive to understand the relations between the complexity classes in the world of communication complexity. Their paper investigates a new 'barrier' in complexity called {\em algebrization}. Without going deeper into this topic one of their results is that a separation between two communication complexity classes shows that the algebrization barrier applies to their Turing machine analogues, i.e., an attempt to show that the two classes are the same will encounter said barrier.

A restricted model of MA (and QMA) one-way communication complexity has been investigated by Aaronson \cite{aaronson:demerlin}.
Very recently Gavinsky and Sherstov \cite{GavinskySherstov} have shown that co-NP is not a subset of MA in the model of multiparty communication complexity.

 One motivation for investigating proof systems in communication complexity is to understand the power of proofs. The results in \cite{raz:am2} show that in a setup where we can actually prove such statements, quantum proofs (of type QMA) are more powerful than classical proofs. However, many questions about this topic remained open: is AM larger than MA? Is co-NP a subset of QMA? Is QMA a subset of AM? In this paper we resolve some of these questions.

In our first result we show a lower bound of $\Omega(n^{1/3})$ for the QMA-communication complexity of $\Disj$.
 This means that co-NP is in fact not a subset of QMA in the world of communication complexity, and that trying to put co-NP into QMA  in the 'real world' (besides the inclusion probably being false) would require a nonalgebrizing technique.

 We also show how a lower bound method we call 'one-sided discrepancy' gives lower bounds for QMA-communication complexity \footnote{Essentially the same method has been used by Gavinsky and Sherstov for the multiparty version of MA-communication complexity \cite{GavinskySherstov}.}. Furthermore we show a (basically tight) lower bound of $\Omega(\sqrt n)$ for the QMA-communication complexity of the function $\IP$, the function that computes the inner product modulo 2. It is an interesting question, whether the bound for $\Disj$ is tight, because we have two very different upper bounds of the order $\widetilde{O}(\sqrt n)$ for this problem: the MA-protocol from \cite{aaronson:alg}, and the Grover based quantum protocol from e.g.~\cite{aaronson&ambainis:search}. Is it possible to combine these protocols in some way to get a more efficient QMA protocol?

Our second main result shows that there is a partial function $f$, for which the weakly unbounded error communication
complexity $PP(f)$ is $\Omega(n^{1/3})$, whereas the AM-communication complexity is only $O(\log n)$. The PP lower bound immediately implies a $\Omega(n^{1/6})$ lower bound for the QMA-communication complexity of the same problem. Hence here the tiniest amount of interaction in classical proof systems (the verifiers may challenge the prover with a public coin message) cannot be simulated by a noninteractive proof system even with the help of quantumness.
In terms of algebrization this result shows that putting AM into PP (or even into MA) needs nonalgebrizing techniques. On the other hand it is widely believed that AM=NP because it might be possible to derandomize AM.

The result has another implication: Since PP-complexity coincides with the {\em discrepancy bound} \cite{klauck:lbqcc}, it turns out to be possible for a function to have polynomial size rectangle covers of its 1-inputs with small error under every distribution on the inputs, yet for some distribution on the inputs every individual rectangle is either exponentially small, or has error exponentially close to 1/2. That means any attempt to prove lower bounds for AM-protocols by considering the properties (error and size) of individual rectangles alone must fail. We believe that it is important to show lower bounds for AM-protocols, because new techniques developed for this problem need to get past this 'rectangle barrier'. Furthermore such a proof would be a first step towards resolving the $\Pi_2\neq \Sigma_2$ problem in communication complexity, one of the biggest problems left open by \cite{bfs:classes}.

Finally, our third result considers the structure of MA-protocols. In general, nondeterministic protocols require no nontrivial interaction between Alice and Bob, i.e., after seeing the proof, Alice can just send one message to Bob, who accepts or rejects. The same is trivially true for AM-protocols (with our definition). Interestingly, and somewhat counterintuitively, Raz and Shpilka show that (within a polynomial increase in communication) one-way communication is also optimal for QMA-protocols. We show that there is a problem, for which one-way MA-communication is exponentially worse than two way randomized communication. This highlights the difference between quantum and classical proofs, and is somewhat reminiscent of the fact, that in the 'real world', quantum proof systems in the class QIP can be parallelized to only 3 rounds \cite{kitaev}, whereas a similar parallelization of classical proofs would collapse the polynomial hierarchy (here parallelization refers to the interaction between the prover and the verifier).

\section{Definitions and Preliminaries}

\subsection{Arthur Merlin Communication Complexity Definitions}

For definitions of more standard modes of communication complexity we refer to Kushilevitz and Nisan's excellent monograph \cite{kushilevitz&nisan:cc}.

In Arthur Merlin communication games, there are 3 parties Merlin, Alice, Bob. All of them are computationally unbounded. Alice sees her input $x\in\{0,1\}^n$, Bob his input
 $y\in\{0,1\}^n$, and Merlin sees both inputs. Merlin is the {\em prover}, who wants to convince the {\em verifier}, consisting of Alice and Bob together, that $f(x,y)=1$.

\begin{Def} In a Merlin-Arthur protocol (short MA-protocol) for a Boolean function $f$ Alice initially receives a message (also called the {\em proof}) from Merlin. After this Alice and Bob communicate until they compute an output, using public key randomness (the proof cannot depend on the randomness). The {\em cost} of an MA-protocol is the sum of the length $a$ of the proof, and the length $c$ of the overall communication between Alice and Bob. The protocol computes $f$, if for all inputs $x,y$ with $f(x,y)=1$ there exists a proof such that $x,y$ is accepted with probability $p$ and for all inputs $x,y$ with $f(x,y)=0$ and all proofs the probability that $x,y$ is accepted is at most $q$. $p$ must be at least a constant factor larger than $q$. $p$ is the {\em completeness}, $q$ the {\em soundness} of the protocols.
We will call $\max\{1-p,q\}$ the {\em error} of the protocol, and frequently consider protocols with very small error.
If not mentioned otherwise we assume $p=2/3$ and $q=1/3$.

The Merlin-Arthur complexity of $f$, denoted $MA(f)$, is the smallest cost of an MA-protocol for $f$.
The MA-complexity with bounded proof length $a$ is denoted $MA^{(a)}(f)$.\end{Def}

Note that the error probabilities (resp.~the soundness and completeness) of MA-protocols can be improved arbitrarily by using standard boosting techniques. For this the proof itself does not need to be repeated, so the proof length is not increased.

Also note that including the proof length in the cost is crucial, because otherwise Merlin could provide Alice with a copy of $y$, whose correctness could be checked with a standard fingerprinting protocol, decreasing the complexity of all functions to $O(1)$ (due to public coins being available).

\begin{Def} In an Arthur Merlin (short AM-) protocol, Merlin, Alice, and Bob share a source of random bits.
 First a random challenge is drawn from this source (of a predefined length). Merlin then produces a message (called the {\em proof}), which is sent to Alice. After this Alice and Bob communicate until they either accept or reject. They may not use fresh random bits at this stage, i.e., all random bits are known to Merlin.
The cost of an AM-protocol is the sum of the length of the proof and the length of the communication between Alice and Bob.

The protocol computes $f$, if for all inputs $x,y$ with $f(x,y)=1$ with probability at least $2/3$ there exists a proof such that $x,y$ is accepted, and for all inputs $x,y$ with $f(x,y)=0$ with probability at most $1/3$ there exists a proof such that $x,y$ is accepted.
The Arthur Merlin complexity of $f$, denoted $AM(f)$, is the smallest cost of an AM-protocol for $f$.
\end{Def}

Note that a more generous definition is possible, in which Alice and Bob still have access to private random bits after receiving the proof. We prefer to call such protocols AMA-protocols, because our definition of AM-protocols is combinatorially cleaner and strong enough for our separation result. Note that by standard techniques from the theory of Arthur Merlin games \cite{babai1,babai2} both MA- and AMA-protocols can be at most quadratically cheaper than AM-protocols, while we later show that AM-protocols can indeed be exponentially more efficient than MA-protocols.

Also note that AM-protocols need only one round of communication between Alice and Bob: to simulate any more complex protocol, Merlin can include the whole conversation between Alice and Bob in his proof, who just need to check if their part was represented properly.

We now define a quantum version of MA-protocols.

\begin{Def} In a quantum Merlin-Arthur (short QMA-)protocol, Merlin, produces a quantum state $\rho$ (the {\em proof}) on some $a$ qubits, which he sends to Alice. Alice and Bob then communicate using a quantum protocol, and either accept or reject the inputs $x,y$. We say that a QMA-protocol computes a Boolean function $f$, if for all inputs $x,y$ with $f(x,y)=1$, there exists a (quantum) proof, such that the protocol accepts with probability at least $p$, and for all inputs $x,y$ with $f(x,y)=0$, and all (quantum) proofs, the protocol accepts with probability at most $q$. Again, we require  $p\gg q$, and we set them to 2/3 resp. 1/3 if not mentioned otherwise. The cost of a QMA-protocol is the sum of $a$ and the length of the communication between Alice and Bob. The cost of the cheapest protocol that computes $f$ defines $QMA(f)$.
The QMA-communication complexity with bounded proof length $a$ is denoted by $QMA^{(a)}(f)$.
\end{Def}

Let us first note that, surprisingly, the error probability of QMA-protocols can be reduced without repeating the quantum proof, due to a clever procedure introduced by Marriott and Watrous \cite{watrous:qma} in the context of standard QMA-games. Since the proof of \cite{watrous:qma} uses the verifier simply as a black box, their construction carries over to the communication complexity scenario. Note however, that this boosting technique increases the number of rounds between Alice and Bob, because their message sequences are computed and uncomputed in a sequential manner.

\begin{Fact}\label{boost}
If there is a QMA-protocol with proof length $a$, communication $c$ and error $1/3$, then there is a QMA-protocol with proof length $a$, communication $O(c\cdot k)$ and error $1/2^k$.
\end{Fact}

We now turn to protocols with (weakly) unbounded error.

\begin{Def} In a weakly unbounded error protocol Alice and Bob have access to a private source of random bits each. The protocol computes a Boolean function $f$ if for all inputs $x,y$ the probability $p_{x,y}$ of computing the correct output $f(x,y)$ exceeds 1/2. The {\em gap} on input $x,y$ is $g_{x,y}=p_{x,y}-1/2$, the gap $g=\min_{x,y} g_{x,y}$. The cost of a weakly unbounded error protocol with worst case communication $c$ is $c-\log g$, and the weakly unbounded error complexity of a function $f$ is $PP(f)$, the minimum cost of any protocol that computes $f$ in the described manner.
\end{Def}

There is another type of unbounded error protocols, in which the gap is not considered, but only the communication necessary to achieve correctness probability exceeding 1/2. We do not consider this model here. See e.g.~\cite{bfs:classes} and \cite{sherstov:ac0, sherstov:pp, buhrman:pp} for more.

\subsection{Integer Polynomials}

In this section we consider the representation of Boolean functions by polynomials with integer coefficients.

Let $f:\{0,1\}^n\to\{0,1\}$ be a Boolean function. A (partial) {\em assignment} $A : S \rightarrow \{0,1\}^m$ is an assignment of values to some subset $S \subseteq \{x_1,\ldots, x_n\}$ of variables. We say that $A$ is {\em consistent} with $x\in\{0,1\}^n$ if $x_i= A(i)$ for all $i\in S$. We write $x\in A$ as shorthand for `$A$ is consistent with $x$'. We write $|A|$ to represent  the cardinality of $S$ (not to be confused with the number of consistent inputs). Furthermore we say that an index $i$ {\em appears} in $A$, iff $i \in S$ where $S$ is the subset of $[n]$ corresponding to $A$. We define a function $\kappa_A:\{0,1\}^n\to\{0,1\}$ such that $\kappa_A(x)=1$ iff $A$ is consistent with $x$

Every $f:\{0,1\}^n\to\{0,1\}$ can be written as $\mbox{sign}(\sum_{A:|A|\leq d} w_A\cdot \kappa_A(x))$, where $d\leq n$ is an integer, the sum is over all partial assignments, and the $w_A$ are integers. We call the minimum of $\sum_{A:|A|\leq d} |w_A|$ that achieves this the {\em threshold weight} $W(f,d)$ of $f$ with {\em degree} $d$. If $d$ is too small to allow representation of $f$, we set $W(f,d)=\infty$. We say that the integer polynomial $\mbox{sign}(\sum_{A:|A|\leq d} w_A\cdot \kappa_A(x))$ {\em sign-represents} the function $f$.

 Frequently in the literature (and importantly for us in Sherstov's paper \cite{sherstov:pattern}) the threshold weight is defined not with partial assignments, but with characters $\chi_S$ of the Fourier transform over the Boolean cube, i.e., parity functions on subsets $S$. Note that this changes the value of $W(f,d)$ at most by a factor of $2^d$: in order to represent the function $\chi_S$ with weight $w_S$ we can assign weight $w_S$ to all partial assignments that fix all variables in $S$ such that the parity of the variables in $S$ is 1, and $-w_S$ to all partial assignments that fix all variables in $S$ such that their parity is 0. Hence we get a representation using partial assignments instead of the $\chi_S$ with total threshold weight increased by a factor of at most $2^d$. Conversely, one can show how given a partial assignment $A$ with weight $w_A$ one can find a representation using a sum of $\chi_S$, so that the overall threshold weight is increased by at most $2^d$. We omit this since it is not important for this paper.

\subsection{Real Polynomials}

In Section 3 we will also use the representation of Boolean functions by polynomials with real polynomials. For definitions concerning this topic we refer to \cite{buhrman:trees}

\subsection{Pattern Matrices}

In \cite{sherstov:pattern} Sherstov introduced a method to turn Boolean functions $f:\{0,1\}^n\to\{0,1\}$ into communication problems that are hard, whenever $f$ is hard under certain measures of complexity. Here we define pattern matrices.

\begin{Def} For a function $f:\{0,1\}^n\to\{0,1\}$ the pattern matrix $P_f$ is the communication matrix of the following problem: Alice receives a bit string $x$ of length $2n$, Bob receives two bit strings $y,z$ of length n each.  The output of the function described by $P_f$ on inputs $x,y,z$ is $f(x(y)\oplus z)$, where $\oplus$ is the bitwise xor, and $x(y)$ denotes the $n$ bit string that contains $x_{2i-y_i}$ in position $i=1,\ldots, n$.
\end{Def}

\section{QMA-complexity of Disjointness}

In this section we prove that the Disjointness problem $\Disj$ requires QMA-communication complexity $\Omega(n^{1/3})$.

Let us first define the problem.

\begin{Def}
The Disjointness problem $\Disj$ has two $n$-bit strings $x,y$ as inputs. $\Disj(x,y)=1\iff \bigwedge_{i=1,\ldots, n} (\neg x_i\vee\neg y_i)$.
\end{Def}

Previous result about this problem are: \cite{bfs:classes} prove $R(\Disj)=\Omega(\sqrt n)$ (and observe that $\Disj$ is complete for the communication complexity version of co-NP). \cite{ks:disj} and later \cite{razborov:disj} prove the tight $\Omega(n)$ bound. \cite{razborov:qdisj} shows the tight $\Omega(\sqrt n)$ lower bound for quantum protocols. This result was reproved in a simpler way in \cite{sherstov:pattern}. \cite{klauck:thresh} gives a $\Omega(\sqrt n)$ lower bound for
MA-protocols. Finally, \cite{aaronson:alg} show a $O(\sqrt n\log n)$ upper bound for MA-protocols, and \cite{aaronson&ambainis:search} give a $O(\sqrt n)$ upper bound for quantum protocols.

Here we prove:

\begin{Thm}\label{Thm:disj}
$QMA(\Disj)=\Omega(n^{1/3})$.
\end{Thm}

We are going to give two proofs of this. The first here uses Razborov's method. Following this (in Section 4) we describe a general method to prove QMA lower bounds, and show how Shertov's technique can be used to yield an overall simpler proof (when taking the proof of Razborov's method into account).

Razborov's method can be summarized as follows \cite{razborov:qdisj}, see also \cite{klauck:qsdpt}.

\begin{Fact}\label{lemrazborov}
Consider a $c$-qubit quantum communication protocol on $n$-bit inputs $x$ and $y$,
with acceptance probabilities denoted by $p(x,y)$.
Define $p(i)=E_{|x|=|y|=n/4, |x\wedge y|=i|}[p(x,y)]$,
where the expectation is taken uniformly over all $x,y$
that each have weight $n/4$ and that have intersection $i$.
For every $d\leq n/4$ there exists a degree-$d$ polynomial $q$ such
that $|p(i)-q(i)|\leq 2^{-d/4+2c}$ for all $i\in\{0,\ldots,n/8\}$.
\end{Fact}

\begin{Proof}[Proof of Theorem \ref{Thm:disj}.]
Suppose we are given a QMA-protocol for $\Disj$ with communication $c$ and proof length $a\geq 1$ and error $1/3$. In what will become a recurring theme, we first amplify the success probability to $1/2^{10a}$ by employing Marriott-Watrous boosting (Fact \ref{boost}). We end up with a protocol that still has proof length $a$, but now the communication is $c'=O(ac)$. We will show that this protocol needs communication at least $\Omega(\sqrt{na})$, which implies the theorem.

At this point we simply replace Merlin's proof with the totally mixed state. We end up with an ordinary quantum protocol, that has the following properties:
\begin{enumerate}
\item All 1-inputs of $\Disj$ are accepted with probability at least $(1-2^{-10a})/2^a$.
\item No 0-input of $\Disj$ is accepted with probability larger than $1/2^{10a}$.
\end{enumerate}

Now we can simply invoke Fact \ref{lemrazborov}. We set $d=12c'$. Then we receive a polynomial $q$, such that
\begin{enumerate}
\item the degree of $q$ is $d$.
\item $1+2^{-c'}\geq q(0)\geq (1-2^{-10a})/2^a-2^{-c'}$.
\item $-2^{-c'}\leq q(i)\leq 2^{-10a}+2^{-c'}$ for all $1\leq i\leq n/8$.
\end{enumerate}

Now we define a rescaled polynomial $q'=1-q/q(0)$.
\begin{enumerate}
\item The degree of $q'$ is $d$.
\item $q'(0)=0$.
\item $1+2^{-c'}/(1+2^{-c'})\geq q'(i)\geq 1-(2^{-10a}+2^{-c'})/((1-2^{-10a})/2^a-2^{-c'})\geq 1-2^{-8a}$ for all $1\leq i\leq n/8$.
\end{enumerate}

The resulting polynomial must rise very steeply between $q'(0)$ and $q'(1)$. We can apply a result by Buhrman et al. \cite{buhrman:qerror}, their Theorem 17.

\begin{Fact}
Every polynomial $s$ of degree $d\leq M-1$ such that $s(0)=0$ and $1-\epsilon\leq s(x)\leq 1$ for all integers $i\in[1,M]$ has
\[\epsilon\geq\frac{1}{u}e^{-v d^2/(M-1)-8d/\sqrt M},\]
where $u,v$ are constants.
\end{Fact}

Setting $M=n/8$, and rescaling $q'$ slightly, we can use this fact to see that $d\geq\Omega(\sqrt{na})$ in order to enable $\epsilon\leq 2^{-\Omega(a)}$.

Hence $c'\geq\Omega(\sqrt{na})$, and $\sqrt ac\geq\Omega(\sqrt{n})$. This implies that $a+c\geq\Omega(n^{1/3})$, which is our theorem.

\end{Proof}

\section{A Lower Bound Method for QMA-protocols}

In this section we develop a general method to prove lower bounds for QMA protocols, and we show how to use the pattern matrix method \cite{sherstov:pattern} for QMA-protocols.

\subsection{A Discrepancy Measure}

Let us start with the familiar notion of the discrepancy bound in communication complexity (see \cite{kushilevitz&nisan:cc}).

\begin{Def}
The (rectangle) discrepancy of a Boolean function $f$ under a distribution $\mu$ is
\[\disc^\mu(f)=\frac{1}{\max_R |\mu(f^{-1}(0)\cap R) -\mu(f^{-1}(1)\cap R)|},\]
where the maximum is over all rectangles $R$ in the communication matrix.

The discrepancy of $f$ is $\disc(f)= \max_\mu\disc^\mu(f)$.\end{Def}

The following linear program (see \cite{jain:partition}) characterizes discrepancy. In the following $\mcR$ denotes the set of all rectangles in the communication matrix.

\vspace{0.1in}

{\footnotesize
\hspace{-0.3in}\begin{minipage}{3in}
    \centerline{\underline{Primal}}\vspace{-0.1in}
    \begin{align*}
      & \text{min}\quad  \sum_{R \in \mcR}  w_{R} + v_R\\
       \quad &  \forall (x,y) \in  f^{-1}(1): \quad \sum_{R: (x,y) \in R} w_{R}  - v_R \geq 1 ,\\
      & \forall (x,y) \in  f^{-1}(0): \quad  \sum_{R: (x,y) \in R} v_R - w_{R} \geq 1,\\
      & \forall R : w_{R}, v_R \geq 0 \enspace .
    \end{align*}
\end{minipage}
\begin{minipage}{3in}
    \centerline{\underline{Dual}}\vspace{-0.1in}
    \begin{align*}
      & \text{max}\quad   \sum_{(x,y)}  \mu_{x,y}  \\
       \quad &   \forall R : \sum_{(x,y)\in f^{-1}(1)\cap R} \mu_{x,y} - \sum_{(x,y)\in R \cap f^{-1}(0) } \mu_{x,y}  \leq 1,\\
      &   \forall R : \sum_{(x,y)\in f^{-1}(0)\cap R} \mu_{x,y}  - \sum_{(x,y)\in R \cap f^{-1}(1) } \mu_{x,y}  \leq 1,\\
      & \forall (x,y) : \mu_{x,y} \geq 0 \enspace .
    \end{align*}
\end{minipage}
}

\vspace{0.1in}

The rectangle discrepancy characterizes the weakly unbounded error communication complexity $PP(f)$, and serves as a lower bound for bounded error quantum and randomized communication. For the bounded error modes it often yields only very poor results. Here is the relation to PP-communication complexity \cite{klauck:lbqcc}.

\begin{Fact}\label{PPdisc}
For all Boolean functions $f:\{0,1\}^n\times \{0,1\}^n\to\{0,1\}$ we have $PP(f)\geq\Omega(\log\disc(f))$ and $PP(f)\leq O(\log\disc(f)+\log n)$.
\end{Fact}

Now we define a lower bound method that we will use for QMA-communication complexity, the one-sided smooth discrepancy. It is similar to the smooth discrepancy \cite{klauck:lbqcc,sherstov:pattern}, in which the primal linear program for discrepancy is augmented with additional upper and lower bounds. Here we augment the program only for the 0-inputs. Essentially the same method (in what we later define as its 'natural' version) was used recently by Gavinsky and Sherstov in the setting of multiparty protocols and MA-communication \cite{GavinskySherstov}.

\begin{Def}[One-Sided Smooth Discrepancy] Let $f:\{0,1\}^n\times\{0,1\}^n \rightarrow \{0,1\}$ be a Boolean function. The one-sided smooth discrepancy of $f$, denoted $\sdisc^1_\epsilon(f)$, is given by the optimal value of the following linear program. \end{Def}

{\footnotesize
\hspace{-0.3in}\begin{minipage}{3in}
    \centerline{\underline{Primal}}\vspace{-0.1in}
    \begin{align*}
      & \text{min}\quad  \sum_{R \in \mcR}  w_{R} + v_R\\
       \quad &  \forall (x,y) \in  f^{-1}(1): \quad\sum_{R: (x,y) \in R} w_{R}  - v_R \geq 1 ,\\
      & \forall (x,y) \in  f^{-1}(0): \quad  1+ \epsilon \geq \sum_{R: (x,y) \in R} v_R - w_{R} \geq 1,\\
      & \forall R : w_{R}, v_R \geq 0 \enspace .
    \end{align*}
\end{minipage}
\begin{minipage}{3in}
    \centerline{\underline{Dual}}\vspace{-0.1in}
    \begin{align*}
      & \text{max}\quad   \sum_{(x,y)}  \mu_{x,y}  - (1+\epsilon)  \phi_{x,y}\\
       \quad &   \forall R : \sum_{(x,y)\in f^{-1}(1)\cap R} \mu_{x,y}  - \sum_{(x,y)\in R \cap f^{-1}(0) } (\mu_{x,y} - \phi_{x,y}) \leq 1,\\
      &   \forall R : \sum_{(x,y)\in f^{-1}(0)\cap R} (\mu_{x,y} - \phi_{x,y}) - \sum_{(x,y)\in R \cap f^{-1}(1) } \mu_{x,y} \leq 1,\\
      & \forall (x,y) : \mu_{x,y} \geq 0 ; \phi_{x,y} \geq 0 \enspace .
    \end{align*}
\end{minipage}
}

\vspace{0.1in}

Note that for all $(x,y)\in f^{-1}(1): \phi_{x,y}=0$ in an optimal solution. We are now looking for a 'natural' definition of one-sided smooth discrepancy, i.e., a definition in which the one-sided smooth discrepancy of a function $f$ is related to the discrepancy of a function $g$ that is similar to $f$. The value of one-sided smooth discrepancy will be the discrepancy of $g$ under a distribution $\nu$. The above dual shows us that we should have $f^{-1}(1)\subseteq g^{-1}(1)$. Furthermore, not too many 0-inputs of $f$ should be 1-inputs of $g$. It is also quite easy to see that for no input $\phi_{x,y}>0$ and $\mu_{x,y}>0$ simultaneously in an optimal solution to the dual.

Below we present the natural definition of one-sided smooth discrepancy.
\begin{Def}[One-sided Smooth Discrepancy, Natural Definition]\label{natural}
Let $f : \{0,1\}^n\times\{0,1\}^n \rightarrow \{0,1\}$ be a Boolean function. The $\delta$-one-sided smooth discrepancy of $f$,  denoted $\tsdisc^1_{\delta}(f) $, is defined as follows:
\begin{align*}
& \tsdisc_{\delta}^1(f) \defeq \max\{\tsdisc^{\lambda,1}_{\delta}(f) : \lambda \mbox{ distribution on } \{0,1\}^n\times\{0,1\}^n \}.\\
& \tsdisc^{\lambda,1}_{\delta}(f) \defeq \max\{\disc^{\lambda}(g) \mbox{ such that } g: \{0,1\}^n\times\{0,1\}^n \rightarrow \{0,1\},\\
& f^{-1}(1)\subseteq g^{-1}(1)\quad \mbox{ and } \lambda(f^{-1}(1))\geq\delta\cdot\lambda(g^{-1}(1))\} .
\end{align*}
\end{Def}

The following lemma shows the equivalence of the two definitions of one-sided smooth discrepancy.
\begin{Lem}\label{lem:sdisceq}
Let $f: \{0,1\}^n\times\{0,1\}^n\rightarrow \{0,1\}$ be a function and let $\delta > 0$. Then
\begin{enumerate}
\item $\tsdisc^1_{\delta/3}(f) \geq \sdisc^1_\delta(f) $.
\item $\delta\cdot\tsdisc^1_{2\delta }(f) \leq \sdisc^1_\delta(f) $.
\end{enumerate}
\end{Lem}

\begin{Proof}
\begin{enumerate}
\item Suppose $\sdisc^1_{\delta }(f)\geq K$. Then there is a solution $\mu,\phi$ to the dual program that achieves value $K$. We have to define a function $g$ and a distribution $\lambda$. W.l.o.g.~we can assume that for every $x,y$ either $\mu_{x,y}=0$ or $\phi_{x,y}=0$. We define $g(x,y)=f(x,y)$ when $\phi_{x,y}=0$, else $g(x,y)=1$. Clearly, $f^{-1}(1)\subseteq g^{-1}(1)$. Furthermore define $L=\sum_{(x,y)}  \mu_{x,y}  +  \phi_{x,y}$. Set $\lambda_{x,y}=(\mu_{x,y}+\phi_{x,y})/L$. The constraints of the dual now imply that $\disc^\lambda(g)\geq L\geq K$.

    Denote $\mu_0=\sum_{x,y:f(x,y)=0}\mu_{x,y}/L,$ and $\mu_1=\sum_{x,y:f(x,y)=1}\mu_{x,y}/L$ and $\phi=\sum_{x,y}\phi_{x,y}/L$.
    Since the set of all inputs is a rectangle, we have that $\mu_0\leq 1/2+1/L$, and $1/2+1/L\geq\mu_1+\phi\geq 1/2-1/L$.

    Assume that $ \lambda(f^{-1}(1))\leq(\delta/3)\cdot\lambda(g^{-1}(1))$. This means that $\mu_1\leq(\delta/3)\cdot(\mu_1+\phi)\leq (\delta/6)+1/L$.
    Then
    \[\mu_1+\mu_0-(1+\delta)\phi\leq \delta/6+1/2-(1+\delta)(1/2-\delta/6)+3/L<0,\]
and the objective function would be negative, which is impossible.

\item Suppose  $\tsdisc^1_{2\delta }(f)\geq K$. Then there are a function $g$ and a distribution $\lambda$ as in the definition of $\tsdisc^1_{2\delta}$.
For all $x,y$ with $g(x,y)=1$ but $f(x,y)=0$ we set $\phi_{x,y}=\lambda_{x,y}\cdot K$, for all other $x,y$ we set $\mu_{x,y}=\lambda_{x,y}\cdot K$. All other variables are set to 0.

Clearly, all constraints of the dual are satisfied. Define  $\mu_0=\sum_{x,y:f(x,y)=0}\mu_{x,y},$ and $\mu_1=\sum_{x,y:f(x,y)=1}\mu_{x,y}$ and $\phi=\sum_{x,y}\phi_{x,y}$.
 Because the set of all inputs is a rectangle, we have that $\mu_0\geq K/2-1$, and $K/2+1\geq\mu_1+\phi\geq K/2-1$.
The objective function is
\[\mu_1+\mu_0-(1+\delta)\phi\geq \delta K-1 + K/2-1 - (1+\delta)(K/2+1-\delta K)\geq \delta K.
\]

\end{enumerate}
\end{Proof}

 We can now show that the one-sided smooth discrepancy yields lower bounds for QMA-communication complexity.

\begin{Thm}\label{QMAdisc}
Let $QMA^{(a)}(f)\leq c$. Then $\log\sdisc^1_{2^{-10a}}(f)\leq O((a+1)c)$.

$QMA(f)\geq\Omega\left(\sqrt{\log\sdisc^1(f)}\right).$
\end{Thm}

One immediate corollary is a lower bound for the function Inner Product mod 2 ($\IP$), because it is well known that even the discrepancy of $\IP$ is at most $2^{-\Omega(n)}$ \cite{chor}.

\begin{Cor}
$QMA(\IP)\geq\Omega(\sqrt{n}).$
\end{Cor}

Note that this lower bound is tight within a log factor due to the MA-protocol of Aaronson and Wigderson \cite{aaronson:alg}.

\begin{Proof}[Proof of Theorem \ref{QMAdisc}.]
Suppose we have a QMA-protocol with proof length $a\geq 1$, communication $c$, and error 1/3, and $a+c$ optimal. We boost the success probability using the Marriott-Watrous technique Fact \ref{boost}. This gives us a QMA-protocol with proof length $a$, communication $c'\leq O(ca)$, and error $2^{-13a}$. We replace the proof at this point with the totally mixed state, leaving us with a quantum protocol that accepts all 1-inputs with probability at least $(1-2^{-13a})/2^a$, and accepts 0-inputs with probability at most $2^{-13a}$.

Our goal is to show, that this gives us a solution to the linear program for one-sided smooth discrepancy. We consider the matrix of acceptance probabilities.

\begin{Def}
  For a matrix $M$ with real entries, denote by $\mu(M)$ the minimum of $\sum_R |u_R|$ (where $R$ ranges over all rectangles in the matrix $M$) such that $M=\sum_R u_R\cdot f_R$, where $f_R$ is the characteristic function of rectangle $R$.
\end{Def}

Linial and Shraibman \cite{linial:norms} have shown the following.

\begin{Fact}
If $M$ is the matrix of acceptance probabilities of a quantum protocol (with shared entanglement) with communication $c$, then $\mu(M)\leq O(2^c)$.
\end{Fact}

So let $M$ be the matrix we constructed before (i.e., for 1-inputs $x,y$ the entry at position $x,y$ is between
$2^{-a}(1-2^{-13a})$ and 1, and for 0-inputs between $0$ and $2^{-13a}$). Consider the system of weights $u_R$ that achieve $\mu(M)\leq O(2^{c'})$.

To turn this into a solution for our primal program for the one-sided smooth discrepancy bound we can simply multiply all the $u_R$ by a factor of $2^{a+2}$ and subtract $1+2^{-12a+2}$ from the $u_R$ for the rectangle $R=\{0,1\}^n\times\{0,1\}^n$. Then, for all $R$, when $u_R<0$ we set $v_R=-u_R$ and $w_R=0$, otherwise we set $w_R=u_R$ and $v_R=0$.
The result is a feasible solution with the parameter $\epsilon\leq 2^{-12a+2}\leq 2^{-10a}$, and the cost of the linear program is at most $O(2^{a}\cdot 2^{c'})\leq O(2^{(a+1)c})$.
Hence  $\log\sdisc^1_{2^{-10a}}(f)\leq O((a+1)c)$.

Also, $QMA(f)\geq a+c\geq\Omega\left(\sqrt{\log\sdisc_{2^{-10a}}^1(f)}\right).$
\end{Proof}

\subsection{Proving Lower Bounds for Pattern Matrices}

In this section we follow the approach of Sherstov \cite{sherstov:pattern}, which can be summarized as follows: for a Boolean function $f:\{0,1\}^n\to\{0,1\}$, we can define a communication problem $P_f$, the pattern matrix (see Section 2.4), and then lower bound the communication complexity of $P_f$ in terms of some parameter of the function $f$. Sherstov uses mainly the approximate degree of $f$ to get lower bounds on the smooth discrepancy of the function $P_f$ (and hence its quantum communication complexity).

Our goal is to relate the one-sided smooth discrepancy of $f$, redefined for query problems, to the one-sided smooth discrepancy of $P_f$. Thanks to the natural definition of $\sdisc^1$ (Definition \ref{natural}) it is actually sufficient to relate the discrepancies of $f$ and $P_f$. In the next section we define discrepancy measures for query complexity.

\subsection{Another Notion of Discrepancy}

In this section we define a notion of discrepancy for Boolean functions (which can be used as a lower bound for query complexity, and in fact characterizes PP-query complexity). Here subcubes defined by partial assignments (see Section 2.2) take the role of the rectangles.

We define the query complexity version of  discrepancy as follows.
\begin{Def}[Discrepancy] Let $f : \{0,1\}^n \rightarrow \{0,1\}$ be a function. The (polynomial) discrepancy of $f$, denoted
$\pdisc(f)$, is given by the optimal value of the following linear program.
\end{Def}

{\footnotesize
\hspace{-0.2in}\begin{minipage}{3in}
    \centerline{\underline{Primal}}\vspace{-0.1in}
    \begin{align*}
      & \text{min}\quad  \sum_{A}  (w_{A} + v_A)\cdot 2^{|A|} \\
       \quad &  \forall x \in  f^{-1}(1): \quad  \sum_{A: x \in A} w_{A}  - v_A \geq 1 ,\\
      & \forall x \in  f^{-1}(0): \quad   \sum_{A: x \in A} v_A - w_{A} \geq 1,\\
      & \forall A: w_{A}, v_A \geq 0 \enspace .
    \end{align*}
\end{minipage}
\begin{minipage}{3in}
    \centerline{\underline{Dual}}\vspace{-0.1in}
    \begin{align*}
      & \text{max}\quad   \sum_{x}  \mu_{x}  \\
       \quad &   \forall A : \sum_{x\in f^{-1}(1)\cap A} (\mu_{x} ) - \sum_{x\in A, x\not\in f^{-1}(1) } (\mu_{x} ) \leq 2^{|A|},\\
      &   \forall A : \sum_{x\in A, x\not\in f^{-1}(1)} (\mu_{x} ) - \sum_{x\in A \cap f^{-1}(1) } (\mu_{x} ) \leq 2^{|A|},\\
      & \forall x : \mu_{x} \geq 0  \enspace .
    \end{align*}
\end{minipage}
}

\vspace{0.1in}

We only give the 'natural' definition of the corresponding notion of one-sided smooth discrepancy. The linear programs are easy to state and analogous to the communication complexity versions.

\begin{Def} Let $f : \{0,1\}^n\to\{0,1\}$ be a Boolean function. The $\delta$-one-sided smooth (polynomial) discrepancy of $f$,  denoted $\spdisc^1_{\delta}(f) $, is defined as follows:
\begin{align*}
& \spdisc_{\delta}^1(f) \defeq \max\{\spdisc^{\lambda,1}_{\delta}(f) : \lambda \mbox{ distribution on } \{0,1\}^n\}.\\
& \spdisc^{\lambda,1}_{\delta}(f) \defeq \max\{\pdisc^{\lambda}(g) \text{ such that } g: \{0,1\}^n \rightarrow \{0,1\}, \\
& f^{-1}(1)\subseteq g^{-1}(1)\quad \mbox{ and } \lambda(f^{-1}(1))\geq\delta\cdot\lambda(g^{-1}(1))\} .
\end{align*}
\end{Def}

Essentially this measure looks at a combination of threshold weight and degree of polynomials that sign-represent a function with a large enough gap, but without requiring integer coefficients.
We now want to relate the query complexity version and the communication complexity version of discrepancy.
Sherstov proved the following statement \cite{sherstov:pattern}.

\begin{Fact}\label{Fact:thresh1}
Let $P_f$ be the pattern matrix of a function $f$, and $d$ a positive integer.

Then \[\disc(P_f)^2\geq\min\left\{\frac{W(f,d-1)}{2^dn},2^{d} \right\},\]
where $W(f,d)$ denotes the threshold weight of $f$ with degree $d$, i.e., the minimum threshold weight of a polynomial with integer coefficients and degree $d$, that sign-represents $f$.
\end{Fact}

Refer to section 2.2 for an explanation of threshold weight as used here (which is slightly different from Sherstov's usage, leading to an extra factor of $2^d$).
Inspecting the proof of Fact \ref{Fact:thresh1} it becomes clear, that the integrality of the polynomial coefficients in the definition of $W(f,d)$
is only used to ensure that the gap between the value of a polynomial on 1-inputs and 0-inputs is at least 1. It turns out we can replace $W(f,d-1)$ by $\pdisc$. Furthermore, since the linear program for $\pdisc$ already incorporates the factors $2^{|A|}$ the minimum with $2^d$ is unnecessary.
An additional advantage of this is that the program for $\pdisc$ (being linear) has a proper dual, whereas Sherstov's proof works with an approximate 'dual' of $W(f,d)$ proved in a combinatorial way (leading to the square on the lhs and the factor $1/n$ on the rhs, see his Theorem 3.4 in \cite{sherstov:pattern}). Modifying his proof this way yields the following theorem.

\begin{Thm}\label{trans}
Let $P_f$ be the pattern matrix of a function $f$.
Then \[\disc^{\mu_\lambda}(P_f)\geq \Omega(\pdisc^\lambda(f) ),\]
where $\mu_\lambda$ is a distribution, in which the inputs $y,z$ to the communication problem $P_f$ (see Section 2.4) are chosen uniformly, and the input bits $x(y)$ are chosen such that $x(y)\oplus z$ is distributed as in $\lambda$. The remaining bits in $x$ are uniform.
\end{Thm}

Thanks to the natural definitions of one-sided smooth discrepancy  we get an analogous result for one-sided smooth discrepancy.
Sherstov's technique allows to transfer a discrepancy lower bound for the function $g$ (from the natural definition of one-sided smooth discrepancy of $f$) to a lower bound on the discrepancy of $P_g$. $P_g$ (together with the distribution $\mu_\lambda$) can then be used as a witness for the hardness of $P_f$.

\begin{Thm}
Let $P_f$ be the pattern matrix of a function $f$.
Then \[\sdisc^1_\delta(P_f)\geq \Omega(\delta\cdot\spdisc_{2\delta}^1(f)) .\]

\end{Thm}

\begin{Proof}
When $\spdisc^1_{2\delta}(f)=K$, then there are a function $g$ and a distribution $\lambda$ as in the definition of $\spdisc^1$.
Then $\disc^{\mu_\lambda}(P_g)\geq\Omega(\pdisc^\lambda(g))$ by Theorem \ref{trans}, and by the definition of $\mu_\lambda$ and the properties of $\lambda,f,g$ we have $\mu_\lambda(P_f^{-1}(1))\geq2\delta\mu_\lambda(P_g^{-1}(1))$.

Then $\tsdisc_{2\delta}(P_f)\geq\Omega(K)$, and consequently  $\sdisc_{\delta}(P_f)\geq\Omega(\delta K)$ with Lemma \ref{lem:sdisceq}.

\end{Proof}

Hence it is enough to analyze the one-sided smooth discrepancy of $f$ to lower bound the QMA-communication complexity of $P_f$.

\subsection{The One-sided Smooth Discrepancy of AND}

The AND function is defined by $AND(x_1,\ldots,x_n)=x_1\wedge\cdots\wedge x_n$. $P_{AND}$  contains the communication matrix of $\Disj$ as a submatrix \cite{sherstov:pattern}.

\begin{Lem}
$\log \spdisc^1_{1/3}(AND)=\Omega(\sqrt n)$.

$\log \spdisc^1_{2^{-a}}(AND)=\Omega\sqrt{an})$.

\end{Lem}

The proof of this lemma is analogous to the corresponding part of the proof of Theorem 1 (and follows from results in \cite{buhrman:qerror}).

\begin{Cor}
  $QMA(P_{AND})=\Omega(n^{1/3})$.
\end{Cor}

\section{AM- vs. PP-communication}

In this section we describe a problem, for which its PP-communication complexity is exponentially larger than both its AM-communication complexity and its co-AM-communication complexity.
 Then we can easily show that also the QMA-communication complexity of the problem must be large. The lower bound for PP-communication complexity also implies that the discrepancy method cannot be applied to get AM-communication complexity lower bounds. This essentially means that to lower bound AM-communication complexity, it is not sufficient to study the properties (size and error) of individual rectangles. Essentially we describe a problem, for which for all distributions on the inputs there is an $O(\log n)$ nondeterministic protocol (i.e., a $\poly(n)$ size cover of the 1-inputs) with constant error, whereas there is a  distribution under which each rectangle has exponentially small discrepancy, i.e., all rectangles are either exponentially small, or they have error exponentially close to 1/2.

\subsection{The Problem}

In \cite{vere:am2} Vereshchagin describes a similar separation for query complexity. We start with his Boolean function, which is a relaxed version of the Minsky-Papert function \cite{minsky}.

\begin{Def}
Let $M$ be a matrix in $\{0,1\}^{n\times m}$. $M$ is {\em good}, if every row of $M$ contains a 1. $M$ is $\delta$-bad, if at least $\delta n$ of its rows contain only zeros.

The function $\AppMP$ takes such matrices $M$ as inputs, accepts good matrices, rejects $\delta$-bad matrices, and is undefined on all other matrices.

We will fix $m=4n^2$ and $\delta=1/2$ in this paper. So the input size of the problem is $N=4n^3$.
\end{Def}

Since the complement of the function $\AppMP$ is not necessarily easy to compute by an AM-query algorithm, Vereshchagin defines a function for which also its complement is easy.

\begin{Def}
  The function $\AppMPC$ takes pairs of Boolean $n\times n$ matrices $M,M'$ as inputs. If $M$ is good, and $M'$ is 2/3-bad, then $\AppMPC(M,M')=1$, and if $M$ is 2/3-bad and $M'$ is good, then $\AppMPC(M,M')=0$. In all other cases the function is undefined.

\end{Def}

We can now state the main result from \cite{vere:am2} in our terminology.

\begin{Fact}\label{Fact:Veresh}
\begin{enumerate}
\item For any polynomial with integer coefficients with degree $d\leq n/2$ that sign-represents the function $\AppMP$,
   the threshold weight is at least $0.5e^{n/15}$, i.e., $W(\AppMP,n/2)\geq 0.5e^{n/15}$.
   \item There is a constant $\zeta$ such that for any polynomial with integer coefficients with degree $\zeta \sqrt n$ that sign-represents $\AppMPC$,
the threshold weight is at least $2^{\zeta n}$, i.e., $W(\AppMPC,\zeta\sqrt  n)\geq 2^{\zeta n}$.
\end{enumerate}
\end{Fact}

We now define communication complexity versions of these problems via pattern matrices.

\begin{Def}
  The function $\PAppMP$ is the communication problem defined by the pattern matrix of the Boolean function $\AppMP$. The function $\PAppMPC$ is the communication problem defined by the pattern matrix of the Boolean function $\AppMPC$.
\end{Def}

We now state the obvious fact that AM-communication is small for $\PAppMPC$ and $\PAppMP$.

\begin{Lem}
\begin{enumerate}
\item  $AM(\PAppMP)=O(\log n)$.
\item $AM(\PAppMPC),AM(\neg \PAppMPC)=O(\log n)$.
\end{enumerate}
\end{Lem}

\begin{Proof}
In an AM-protocol for $\PAppMP$, the public coin random number $i$ represents a random row of the matrix $M$ which is the input to the function $\AppMP$ encoded in the pattern matrix. Merlin replies with a position $j$ such that $M(i,j)=1$, if such a $j$ exists. Alice and Bob  can easily verify with logarithmic communication whether $M(i,j)=1$. They accept if this is the case. Clearly, if the $i$-th row of $M$ does not contain a 1 they will not accept any proof. On $\delta$-bad matrices this happens with probability at least $\delta$. Good matrices $M$ on the other hand have their inputs to their communication problem accepted with probability 1.
The remaining protocols are along the same lines.
\end{Proof}

We now turn to the lower bound. As it happens, all we really need to do is to (again) appeal to a result by Sherstov \cite{sherstov:pattern} (restated here for convenience).

\begin{Fact}\label{Fact:thresh}
Let $P_f$ be the pattern matrix of a function $f$, and $d$ a positive integer.

Then \[\disc(P_f)^2\geq\min\left\{\frac{W(f,d-1)}{2^dn},2^{d} \right\},\]
where $W(f,d)$ denotes the threshold weight of $f$ with degree $d$, i.e., the minimum threshold weight of a polynomial with integer coefficients and degree $d$,
that sign-represents $f$.
\end{Fact}

See section 2.2 for an explanation of threshold weight. Putting these results together we find that for $d= n/100$ the function $\AppMP$ has threshold weight at least $2^{\Omega(n)}$, and then the pattern matrix has
discrepancy at least $2^{\Omega(n)}$. This readily implies that $PP(\PAppMP)=\Omega(n)$ with Fact \ref{PPdisc}. Similarly, choosing $d=\zeta\sqrt n$ we get $PP(\PAppMPC)\ge\Omega(\sqrt n)$.

\begin{Thm}
\begin{enumerate}
\item  $PP(\PAppMP)=\Omega(N^{1/3})$.
\item  $PP(\PAppMPC)=\Omega(N^{1/4})$.
\item  $AM(\PAppMP)=O(\log N)$.
\item $AM(\PAppMPC),AM(\neg\PAppMPC)=O(\log N)$.
\end{enumerate}
\end{Thm}

\subsection{QMA vs. AM}

In this subsection we note the following consequence of the lower bound in the previous subsection, which follows from the fact that PP-protocols can simulate QMA-protocols within a quadratic increase in communication (this can be proved by first boosting with Fact \ref{boost}, then removing the proof, which leaves a quantum protocol with a large enough gap. This can be turned into a weakly unbounded error quantum protocol, and such protocols are exactly as powerful as classical weakly unbounded error protocols \cite{klauck:lbqcc}).

\begin{Cor}
\begin{enumerate}
\item  $QMA(\PAppMP)=\Omega(N^{1/6})$.
\item  $AM(\PAppMP)=O\log N)$.
\end{enumerate}
\end{Cor}

\section{Rounds in MA-communication}

For many 'realistic' modes of communication complexity there are problems that require the players Alice and Bob to interact by using many rounds of communication in order to achieve good protocols. This is not altogether surprising, since one would expect conversations with many rounds of interaction to be more powerful than monologues. Examples of this phenomenon are deterministic, randomized (see \cite{nisan:rounds}), and quantum communication complexity (see \cite{klauck:rounds,jain:pointer}).

However, in the nondeterministic mode of communication, monologues are in fact optimal: the prover can provide the whole conversation to the players, who now just need to verify that their role in the conversation is represented correctly.

In this section we show that there is a partial function, for which every one-way MA-protocol with communication going from Alice to Bob is exponentially more expensive
than a randomized one-way protocol with communication going from Bob to Alice. Note that such problems trivially do not exist when we replace MA- with nondeterministic, or AM-communication complexity. Raz and Shpilka \cite{raz:am2} prove that one-way communication (in any direction) is also optimal (within a polynomial increase in communication) for QMA protocols\footnote{They prove that every problem with QMA-communication complexity $c$ can be reduced to a problem called $LSD$ of size $2^{\poly\log c}$, for which a logarithmic QMA one-way protocol exists (they call this a two round protocol).}. Rounds in MA-communication do not seem to have been considered before, although Aaronson \cite{aaronson:demerlin} considers a weaker variant of one-way MA-protocols: Merlin sends the proof to Bob only, so that Alice has to send her message without having seen the proof. This model is much weaker than the standard one-way MA-communication model, in fact at most quadratically more efficient than randomized one-way communication.

Let us define the function. A usual suspect for this kind of separation is the Index function $\Ix$, for which Alice receives a string $x\in\{0,1\}^n$, and Bob
 an index $i\in\{1,\ldots, n\}$ and the goal is to compute $x_i$ (see \cite{knr:index}). But due to Bob's input being short, the nondeterministic complexity of this problem is small, and hence also the one-way MA-communication: The prover can simply provide Alice with Bob's input $i$. The problem we use instead gives Bob many indices, and we are trying to determine whether for many of them $x_i=1$.

 \begin{Def}
   The function $\MajIx(x,I)$, where $I=\{i_1,\ldots, i_{\sqrt n}\}$, each $i_j\in\{1,\ldots,n\}$, and $x\in\{0,1\}^n$ is defined as follows:
   \begin{enumerate}
     \item if $|\{j:x_{i_j}=1\}|=\sqrt n$ then $\MajIx(x,I)=1$,
     \item if $|\{j:x_{i_j}=1\}|\leq 0.9\sqrt n$ then $\MajIx(x,I)=0$,
    \item  otherwise $\MajIx(x,I)$ is undefined.
   \end{enumerate}
 \end{Def}

It is easy to see that $R^{B\to A}(\MajIx)=O(\log n)$, because Bob can just pick $100$ indices $i_j$ from $I$ randomly, and send them to Alice, who accepts if and only if all $x_{i_j}=1$. In AM-protocols a single round of communication from Alice to Bob is always optimal (and finding such a protocol for $\MajIx$ is an easy exercise).

Intuitively, Merlin's problem with $\MajIx$ is that he cannot provide information about many indices in $I$, unless his proof is very long. However, it is not clear at all that this is necessary, and indeed, the same intuition would apply to the quantum case, in which there is a one-way protocol with $\poly\log(n)$ communication and proof length  for the problem. We prove the following lower bound, which is close to optimal.

\begin{Thm}
$MA^{A\to B}(\MajIx)\geq\Omega(\sqrt n).$
\end{Thm}

\begin{Proof}
Before starting let us define the notion of a one-way rectangle, which is a basic object when considering randomized one-way communication complexity.

\begin{Def}
  For a (partial) function $f:X\times Y\to \{0,1\}$ a {\em one-way rectangle} is a subset $R\subseteq X$, coupled with a function $d_R:Y\to\{0,1,-1\}.$ The one-way rectangle {\em accepts} all inputs
$(x,y)\in R\times Y$ with $d_R(y)=1$, {\em rejects} all inputs $(x,y)\in R\times Y$ with $d_R(y)=0$, and is undecided about the remaining inputs in $R\times Y$. The {\em error} of a one-way rectangle under some distribution is defined in the obvious way.
\end{Def}

The size of small error one-way rectangles under distributions on the inputs characterizes the one-way randomized communication complexity \cite{klauck:qcst},\cite{jain:subd}.

Now let us begin with the proof.
  We are given an $MA^{A\to B}$ protocol, which has proof length at most $a$, and communication at most $c$, and error 1/3. We assume that $a,c\leq\gamma\sqrt n$ for some small constant $\gamma$. As usual we boost the success probability by repeating the communication among Alice and Bob $100\sqrt n$ times, so that the (soundness and completeness) error drops to $\epsilon=2^{-10\sqrt n}$.
  Note that the proof does not need to be repeated, so after this step the proof length is still $a$, and the communication is $c'\leq \delta n$ for some small constant $\delta$.
  \footnote{In the quantum case this appears to be impossible, because the Marriott-Watrous boosting technique does not work for one-way protocols. Hence here is a point where a $QMA^{A\to B}$ lower bound along these lines would fail.}

Our argument can now be summarized as follows: First we consider a distribution $\mu$ on 1-inputs. We find and fix a proof for which many 1-inputs are accepted with high probability (we will identify proofs with the sets of 1-inputs that are accepted  with high probability when using those proofs). After fixing the proof we are left with a randomized one-way protocol, that accepts all 1-inputs in the proof with probability $1-\epsilon$, but accepts 0-inputs with probability at most $\epsilon$ each. Now we define a distribution $\sigma$ on 0-inputs. Finally, we show that under the  distribution, in which 1-inputs are chosen according to $\mu$ and 0-inputs according to $\sigma$ any large one-way rectangle must have large error. This shows that $c'=\Omega(n)$, and hence $a+c=\Omega(\sqrt n)$.

So let us begin with the distribution $\mu$ on 1-inputs. For this we employ a good error-correcting code to generate $x\in\{0,1\}^n$. It does not matter whether the code is constructible or not, so a randomized construction suffices. A simple modification of the Gilbert-Varshamov bound gives us a code $C\subseteq\{0,1\}^n$ that has distance $n/4$ and $2^{0.87n}$ codewords, each of which has Hamming weight exactly $n/2$.

For the distribution $\mu$ we first uniformly choose an element $x\in C$ from the code. Denote by ${\cal I}$ the set of all sets $I$ of $\sqrt n$ different indices from $\{1,\ldots, n\}$.
We continue by choosing an index set $I=\{i_1,\ldots, i_{\sqrt{n}}\}\in\cal I$, under the condition that all of the $i_j\in I$ satisfy $x_{i_j}=1$. This finishes the description of $\mu$.

In our MA-protocol there are at most $2^a$ different proofs $p$. We identify each proof $p$ with the set of 1-inputs $(x,I)$, for which Alice and Bob accept $(x,I)$ with probability at least $1-\epsilon$ when given proof $p$. Since completeness is $1-\epsilon$ every 1-input is in at least one proof. No 0-input is in any proof. Now we simply fix the largest proof $p$ under the distribution $\mu$. Then $\mu(p)\geq 2^{-a}$.

Having fixed our proof $p$, we are left with a randomized one-way protocol that accepts at least the 1-inputs in $p$ with probability $1-\epsilon$, and accepts no 0-input with probability larger than $\epsilon$. In order to show that this protocol needs communication $\Omega(n)$ we create a hard distribution on all inputs, by mixing $\mu$ with a distribution on 0-inputs: this distribution $\sigma$ is simply uniform on all 0-inputs.

Note that there are more 0-inputs than 1-inputs to the function $\MajIx$, due to the promise definition. However, when we denote the total number of 1-inputs $(x,I)$ with $x\in C$ by $K$
and the total number of 0-inputs $(x,I)$ with $x\in C$ by $L$, then a simple calculation reveals that $L/K\leq 3^{\sqrt n}$.
We can conclude that for each 0-input $(x,I)$ and each 1-input $(x',I')\in p$ we have
\begin{equation}\label{eq:musigma}
\mu(x',I')\leq 3^{\sqrt n}\cdot\sigma(x,I).\end{equation}

Let $\nu$ be the distribution on all inputs, which results from mixing $\mu$ and $\sigma$ with probability 1/2 each. We may now fix the remaining randomness in our protocol and get a deterministic one-way protocol, that has communication $c'$, and under $\nu$ accepts a set $p'\subseteq p$ of 1-inputs with $\nu(p')\geq 2^{-a}\cdot(1-\epsilon)/2$, but accepts a set $q$ of 0-inputs with $\nu(q)\leq \epsilon/2$. Note that $\epsilon \ll 2^{-a}(1-\epsilon)$.

To simplify our argument we will remove the 1-inputs that have limited contribution to the size of $p'$. A string $x\in C$ is {\it slim} if
\[\frac{\nu(\{(x,I): (x,I)\in p'\})}{\nu(\{x\}\times{ \cal I}    )}  \leq 2^{-2a}.\]
 Let $p''$ denote the set of inputs $(x,I)\in p'$ such that $x$ is not slim. Then $\nu(p'')\geq 2^{-a}\cdot(1-\epsilon)/2-2^{2a}\geq 2^{-a-2}$.

Our goal is to show that under the distribution $\nu$ every one-way protocol with the above properties must have communication $\Omega(n)$. A one-way protocol partitions the inputs into one-way rectangles. Let us consider a one-way rectangle $R,d_R$, such that there are $x\neq y$ with $x,y\in C\cap R$, and both $x,y$ are not slim. Recall that $x$ and $y$ have Hamming distance $n/4$.

\begin{Lem}\label{LemRounds}
If $x,y$ are both in the same one-way rectangle, then the error (restricted to rows that are not slim) is
\[\frac{\nu((R\times d^{-1}_R(1))\cap \MajIx^{-1}(0))}{\nu(R\times{ \cal I}    )}   \geq 2^{-2\sqrt n }.\]
\end{Lem}

Since the protocol accepts only a set of 0-inputs with size at most $\epsilon/2$ under $\nu$, and accepts all inputs in $p''$, at most half of all inputs in $p''$ (under $\nu$) can be in one-way rectangles that contain at least 2 different codewords as rows, else the error exceeds $2^{-a-3}\cdot 2^{-2\sqrt n}\gg\epsilon/2$.

This means that there must be $2^{\Omega(n)}$ one-way rectangles in the protocol, and hence the communication $c'\geq\Omega(n)$, which in turn implies $c+a\geq\Omega(\sqrt n)$, finishing our proof.

\end{Proof}

\begin{Proof}[Proof of Lemma \ref{LemRounds}.]
Consider a one-way rectangle $R,d_R$. $R$ contains at least two different codewords $x\in C$ and $y\in C$ (that are both not slim). We are interested in the restrictions that this places on $d_R$. We identify $x,y$ with subsets of $\{0,1\}^n$. Then $x\cap y\leq n(1/2-1/8)$ because $x$ and $y$ have Hamming distance at least $n/4$. Let $S\subseteq\cal I$ be the set of $I$ such that $(x,I)\in p''$. Then for all $I\in S$ we have that all of the $i\in I$ must satisfy $x_i=1$.

First let $T\subseteq S$ be the set of all $I\in S$, such that $|I\cap x\cap y|\geq 0.8\sqrt n$. Then
\[|T|\leq \sqrt n\cdot{\frac{3n}{8}\choose {0.8\sqrt n}}\cdot {\frac{n}{8}\choose {0.2\sqrt n}}   \leq 2^{-\alpha \sqrt n}\cdot{\frac{n}{2}\choose {\sqrt n}}   ,  \]
for some  constant $\alpha>0$. Note that the binomial coefficient on the right hand side is just the number of $I\in\cal I$ such that $(x,I)$ is a 1-input. Hence
\[\frac{\nu((\{x\}\times T)\cap p'')}{\sum_{I\in {\cal I}: \MajIx(x,I)=1} \nu(x, I)}\leq 2^{-\alpha \sqrt n}.\]
Since we can limit $a$ such that $2^{-2a}\geq 2\cdot2^{-\alpha\sqrt n}$ the set $T$ contributes little to the set of $I\in \cal I$ with  $(x,I)\in p''$:
\[\nu(\{(x,I): (x,I)\in p''\mbox{ and } I\not\in T\})\geq 2^{-2a-1}\cdot \sum_{I\in {\cal I}: \MajIx(x,I)=1} \nu(x, I)\geq 2^{-2a-2}\cdot\nu(\{x\}\times\cal I). \]

So let us examine the set  of all $I\in S$ such that $|I\cap x\cap y|< 0.8\sqrt n$. For all such $I$ we have either  $|x\cap I|\leq .9\sqrt n$ or $|y\cap I|\leq .9\sqrt n$, i.e., either $(x,I)$ or $(y,I)$ is a 0-input. Since we have assumed that $(x,I)\in p''$, we get that $(y,I)$ is a 0-input.  Now 0-inputs have a smaller probability each than 1-inputs, but on the other hands the allowed error $\epsilon$ is very small.

We can now calculate the error of the one-way rectangle in the rows $x$ and $y$.
\begin{eqnarray*}
&&\nu(\{(y,I):\MajIx(y,I)=0\mbox{ and }d_R(I)=1\})\\
 \geq&&
 \frac{\nu(\{(x,I):  I\in S-T    \})}  {3^{\sqrt n}}\\
 \geq&& 2^{-2a-2}/3^{\sqrt n}\cdot\nu(\{x\}\times \cal I\}).\end{eqnarray*}

We can play this game also with $x$ and $y$ exchanged and hence the two rows $x,y$ together have substantial error. We can continue with more pairs of rows $x,y$ of the one-way rectangle, until only one (or no) row is left. Hence the overall error is at least $2^{-2\sqrt n}$.

\end{Proof}

The proof technique used above can also be used to give a simple proof of a good lower bound for the randomized one-way communication complexity of the Index function $\Ix$. For an error parameter $\epsilon$ choose a binary code with distance $\epsilon$ and size $2^{(1-H(\epsilon)n}$. The hard distribution is the uniform distribution on the code times the uniform distribution on Bob's inputs. Whenever two codewords $x,y$ are in the same message, the error in their rows together must be at least $\epsilon$. Hence most codewords must be in separate messages.

\bibliographystyle{alpha}

\newcommand{\etalchar}[1]{$^{#1}$}

\end{document}